\documentclass[12pt]{article}
\usepackage{graphicx}
\usepackage{epsfig}
\usepackage{cite}

\def\href#1#2{#2}

\def\beq{\begin{equation}}
\def\eeq{\end{equation}}

\begin{document}


\renewcommand{\thefootnote}{\fnsymbol{footnote}}
\newcommand{\da}{\dot{a}}
\newcommand{\db}{\dot{b}}
\newcommand{\dn}{\dot{n}}
\newcommand{\dda}{\ddot{a}}
\newcommand{\ddb}{\ddot{b}}
\newcommand{\ddn}{\ddot{n}}
\newcommand{\pa}{a^{\prime}}
\newcommand{\pb}{b^{\prime}}
\newcommand{\pn}{n^{\prime}}
\newcommand{\ppa}{a^{\prime \prime}}
\newcommand{\ppb}{b^{\prime \prime}}
\newcommand{\ppn}{n^{\prime \prime}}
\newcommand{\fda}{\frac{\da}{a}}
\newcommand{\fdb}{\frac{\db}{b}}
\newcommand{\fdn}{\frac{\dn}{n}}
\newcommand{\fdda}{\frac{\dda}{a}}
\newcommand{\fddb}{\frac{\ddb}{b}}
\newcommand{\fddn}{\frac{\ddn}{n}}
\newcommand{\fpa}{\frac{\pa}{a}}
\newcommand{\fpb}{\frac{\pb}{b}}
\newcommand{\fpn}{\frac{\pn}{n}}
\newcommand{\fppa}{\frac{\ppa}{a}}
\newcommand{\fppb}{\frac{\ppb}{b}}
\newcommand{\fppn}{\frac{\ppn}{n}}
\newcommand{\A}{A}
\newcommand{\B}{B}
\newcommand{\mmu}{\mu}
\newcommand{\mnu}{\nu}
\newcommand{\ii}{i}
\newcommand{\jj}{j}
\newcommand{\jl}{[}
\newcommand{\jr}{]}
\newcommand{\ml}{\sharp}
\newcommand{\mr}{\sharp}
\newcommand{\N}{{\cal N}} 
\newcommand{\F}{{\cal F}} 

\title{{\bf Notes on Properties of Holographic Matter}}
\author{A.~Karch$^1$\footnote{email: {\tt karch@phys.washington.edu}}, \,
M.~Kulaxizi$^2$\footnote{email: {\tt M.Kulaxizi@uva.nl}}, \, and A.~Parnachev$^3$\footnote{email: {\tt andreipv@max2.physics.sunysb.edu}}
\\ \vspace{6pt}
{\small\em $^1$Department of Physics, University of Washington, Seattle, WA 98195-1560 USA}
\\ {\small\em$^2$ITFA, University of Amsterdam, Valckenierstraat 65, 1018XE Amsterdam, The Netherlands}
\\ {\small\em$^3$YITP, Stony Brook University, Stony Brook, NY 11794-3840 USA}
}
\vspace{8pt}
\date{August 2008}
\maketitle

\begin{abstract}
\noindent Probe branes with finite worldvolume electric flux in
the background created by a stack of Dp branes describe holographically
strongly interacting fundamental matter at finite density.
We identify two quantities whose leading low temperature behavior
is independent of the dimensionality of the probe branes:
specific heat and DC conductivity.
This behavior can be inferred from the dynamics of the fundamental strings
which provide a good description of the probe branes in the regime
of low temperatures and finite densities.
We also comment on the speed of sound on the branes and the temperature
dependence of DC conductivity at vanishing charge density.

\end{abstract}
\begin{flushleft}
\end{flushleft}
\vspace{-6in}

\newpage

\section{Introduction and Results}

Over the last few years several interesting field theories at finite density have been analyzed using probe D-branes \cite{Karch:2000gx,Karch:2002sh}
in the gauge gravity correspondence \cite{Maldacena:1997re,Witten:1998qj,Gubser:1998bc}. Typically one starts with the supergravity background corresponding to a black Dp brane. The dual field theory is maximally supersymmetric $SU(N_c)$ Yang-Mills (SYM) theory in p+1 dimensions \cite{Itzhaki:1998dd}. In the special case of $p=3$ the gauge theory, ${\cal N}=4$ SYM, is a conformal field theory, but for other values of $p$ the gauge coupling is dimensionfull, so the gauge theory has an intrinsic scale set by the 't Hooft coupling constant $\lambda = g^2_{YM} N_c$. In this background one introduces matter by adding additional fields in the fundamental representation of the $SU(N_c)$ gauge group. In the gravity dual the dynamics of these extra matter fields is captured by the introduction of a probe Dq brane. In the limit of large $N_c$ loops of the fundamental fields can be neglected, as they give
subleading (in $N_c$) effects. The fundamental matter simply experiences dynamics mediated by the strong interactions
of the background SYM. Correspondingly in the gravitational description the probe brane in this limit does not backreact on the geometry. Many different choices of probe branes are possible. The localized matter can live in all of
the $p+1$ dimensions of the background SYM, but it can also occupy lower dimensional defects. We can add supersymmetric matter with an equal amount of fermions and bosons (e.g. whenever the number of relative ND directions between Dp and Dq brane is 4) or fermions only (when the relative number of ND directions is 6). The most studied systems of this type are probably the D3/D7 system of \cite{Karch:2002sh} that was extensively studied at finite density in \cite{Kobayashi:2006sb,Mateos:2007vc,Karch:2007br,Ghoroku:2007re,Nakamura:2007nx,
Erdmenger:2007ja,Erdmenger:2008yj} and
most recently in \cite{Faulkner:2008hm}, the D3/D5 system of \cite{Karch:2000gx}, the D4/D6 system \cite{Kruczenski:2003uq} studied at finite density in \cite{Matsuura:2007zx} and the Sakai-Sugimoto model \cite{Sakai:2004cn} (or rather a simplified version of it with a single chiral multiplet on a 3+1 dimensional defect coupled to uncompactified supersymmetric 4+1 SYM) studied at finite density in \cite{Kim:2006gp,Horigome:2006xu,Parnachev:2006ev,Davis:2007ka,Bergman:2007wp,Rozali:2007rx,Kulaxizi:2008jx}. But of course there are many more options, basically any combination of Dp and Dq system gives a new, different holographic matter system.

In order to organize what is known about the properties of these holographic matter systems, we identify two interesting properties that are q-independent, that is they only depend on the dimensionality of the background Dp brane and are insensitive to the details of the probe. We will mostly focus on the case where the matter multiplets we added are massless. Despite this, it is straightforward to show that two of the basic material properties, the {\it heat capacity} and the {\it DC conductivity}, are completely determined (at large density, small temperature) by the properties of a single heavy quark, represented in the bulk as a semi-classical string ending on the brane. As the properties of the string are completely independent of what Dq probe brane they eventually end on, this equivalence makes it clear that heat capacity and conductivity of the probe brane can not depend on q either.

\begin{table}
\label{table}
\begin{center}
\begin{tabular}{l||l|l|l}
p & Free energy & Heat capacity & Resistivity\\
\hline \hline
0& $T^{2/5}$ & $T^{-3/5}$ & no electric field possible \\
1& $T^{1/4}$ & $T^{-3/4}$ & $T^{3/2}$ \\
2& $T^{2/3}$ & $T^{-1/3}$ & $T^{5/3}$ \\
3& $T$ & q-dependent & $T^{2}$ \\
4& $T^2$ & $T$ & $T^{3}$ \\
\end{tabular}
\end{center}
\caption{Leading density dependent contribution to the
free energy, heat capacity and resistivity for Dq probe branes in a Dp background. This scaling is universal
in that it only depends on the p of the background brane and is independent of the q of the probe brane. The only
exception is the heat capacity in the case of a background D3 brane, where the universal linear in $T$ contribution
to the free energy does not contribute to the heat capacity (rather it gives a finite entropy density at zero
temperature). So the subleading term in the free energy (which is q-dependent) becomes the leading term in the
heat capacity.}
\end{table}

In table \ref{table} we summarized our main results for the heat capacity and conductivity. While some of these have appeared in the literature before (in particular the D3/D7 system was analyzed in \cite{Karch:2008fa} and the D4/D8 system in \cite{Kulaxizi:2008jx}) relating it to the properties of a single string and noting the q-independence allows us to give a answer for the generic Dp/Dq system in this compact form. If one is looking for a system with fermions only, the linear (in $T$) heat capacity of the D8 probe in the D4 background looks encouraging \cite{Kulaxizi:2008jx}. In this case the only matter we added were fundamental fermions, so one may take the linear heat capacity as a strong hint that the fermions indeed dominate the finite density physics. One caveat one should note here is that, while the only matter we added that is explicitly charged under the global $U(1)_B$ for which we turn on a chemical potential are the fundamental fermions, there are
instantonic configurations in the 5d SYM that can also carry $U(1)$ charge. This is the well known anomaly-inflow mechanism of \cite{Callan:1984sa}. As the linear heat capacity is universal to all probes of the D4 background, including for example the supersymmetric D6 brane of \cite{Kruczenski:2003uq} that introduces bosons as well as fermions, it is clear that more studies are needed to settle the microscopic nature of the properties one finds.

The organization of this paper is as follows. In the next section we review the thermodynamics of Dp branes and introduce the corresponding background geometry. In section 3 we calculate the free energy of a single string in the general Dp background and show that the heat capacity of any Dq probe, to leading order in the density, is just this free energy of a single string times the density. We also obtain from this the speed of sound which, unlike the heat capacity, depends on the details of the probe. In section 4 we generalize this to the massive case
and show that the DBI receives the relevant contribution at low temperatures from fundamental strings dissolved in the Dq probe. In section 5 we calculate the drag experienced by a single string and show that this directly gives the DC conductivity of the generic Dq probe brane. In section 6 we turn to the DC conductivity of Dq probe matter at zero density. This quantity is no longer q-independent but depends on details of the probe. We briefly comment on cases in which one has a resistivity linear in temperature. We conclude in section 7. Our main results are summarized in Table 1 (the universal heat capacity and conductivity at temperatures much less than the scale set by the charge density) and Figures 1 and 2 (the non-universal speed of sound and zero density conductivity).

\section{Thermodynamics of the Dp brane}

The thermodynamics of the general Dp brane appeared e.g. in
\cite{Itzhaki:1998dd,Peet:1998wn,Mateos:2007vn} where we will mostly follow the conventions of the last reference. The near horizon geometry of a stack of non-extremal $N_c$ Dp branes is described by the following metric, dilaton and RR form fields:
\begin{eqnarray*}
ds^2 &=& H^{-1/2} ( -f dt^2 + dx_p^2) + H^{1/2} \left ( \frac{du^2}{f} + u^2 d \Omega_{8-p}^2 \right ) \\
e^{\Phi} &=& H^{\frac{3-p}{4}}, \,\,\,\,\,\,\,\,\,\, C_{01\ldots p} = H^{-1}
\end{eqnarray*}
where $H(u) = (L/u)^{7-p}$ and $f(u)=1-(u_h/u)^{7-p}$. It is convenient to work in units in which the curvature radius of the supergravity solution is $L=1$, so that the string length becomes
\beq
l_s^{p-7} = g_s N_c (4 \pi)^{\frac{5-p}{2}} \Gamma(\frac{7-p}{2})
\eeq
or better expressed in terms of the dimensionfull Yang-Mills coupling
$g^2_{YM} = 2 \pi g_s (2 \pi l_s)^{p-3}$ with $\lambda=g^2_{YM} N_c$ as
\beq  l_s^{2(p-5)} = 2^{7-2p} \pi^{\frac{9-3p}{2}} \Gamma(\frac{7-p}{2}) \, \lambda\eeq
The temperature $T$ is given in terms of $u_h$ via
$$ u_h = \left ( \frac{4 \pi T}{7-p}\right )^{2/(5-p)}$$

In these units the supergravity on-shell action, which is minus the free energy density $\omega$ times the volume of the field theory spacetime, is independent of $\lambda$ except for
the overall prefactor, so the free energy scales as $\frac{1}{g_s^2 l_s^8} \sim \lambda^{\frac{p-3}{5-p}}$. The temperature dependence then by dimensional analysis has to be $T^n =T^{\frac{ 2 (7-p) }{5-p}}$. Including order 1 numbers one has
\beq
\frac{\omega}{N^2} = -\frac{1}{n-1} \frac{\epsilon}{N^2} =
-\frac{1}{n} \frac{s T}{N^2} = (p-5) \left ( 2^{29-5p} (7-p)^{-3 (7-p)}
\pi^{13-3p} \Gamma \left ( \frac{9-p}{2} \right )^2 \right )^{\frac{1}{5-p}} \, \lambda^{\frac{p-3}{(5-p)}} T^n
\eeq
where $\epsilon$, $f$ and $s$ are the energy-, free energy- and entropy-density respectively. Correspondingly the speed of sound is $c_s^2 = 1/(n-1) = (5-p)/(9-p)$.

\section{Free energy of a single quark and the Heat Capacity of a Dq probe}

The shift in the free energy of a single quark (often somewhat loosely referred to as the mass shift) due to the horizon is simply given by the change in length
of the string, $u_h$, times the tension of the quark, $1/(2 \pi l_s^2)$, so
\beq
\Delta m = \frac{u_h}{2 \pi \alpha'} = \left ( 2^{6-p} \pi^{(3-p)/2} \Gamma \left ( \frac{7-p}{2} \right )
(7-p)^{-2} \right )^{\frac{1}{5-p}} \lambda^{\frac{1}{5-p}} T^{\frac{2}{5-p}}
\eeq
The shift in free energy is negative.
For $p=3$ this reduces to the result $\Delta m = \frac{ \sqrt{2 \lambda}}{2} T$ of \cite{Herzog:2006gh}\footnote{The normalization for $g^2_{YM}$ we are using here follows \cite{Itzhaki:1998dd,Mateos:2007vn} and differs by a factor
of 2 from \cite{Herzog:2006gh} so that in that work $\tilde{\lambda} = 2 \lambda$. As explained in
\cite{Myers:2007we}
this difference can be traced through to a different convention used for the $U(N_c)$ generators in the literature on D-branes (where $Tr (T_a T_b) = \delta_{ab}$) and gauge theories (where typically $Tr (T_a T_b) = \frac{1}{2} \delta_{ab}$).}. From this one can get the entropy of the single quark as $S=\partial (\Delta m)/\partial T$ and similarly the contribution to the heat capacity from a single quark. We want to show that the free energy of a probe Dq brane describing massless flavors, potentially localized on a defect, is simply $d \, \Delta m$, where $d$ is the number density of quarks (which we refer to as baryon number density, even though we are
taking the convention that a single quark carries $U(1)_B$ charge 1).

To calculate the heat capacity of the probe brane directly, one starts with the Dirac-Born-Infeld (DBI) action describing the worldvolume fluctuations of the brane.
Here and in the rest of the paper we only consider black hole embeddings,
that is embeddings of the brane in which the worldvolume crosses the black
hole horizon. If the brane ends outside the horizon an explicit
source of $U(1)_B$ charge has to be introduced and the properties
of the brane depend crucially on the properties of this object.
For massless flavors the embedding
(in some frame) always has the form AdS$_{d_s+2}$ $\times$ S$^{q - d_s -1}$. Here $d_s$ denotes the number of spatial dimensions of the defect on which the flavors associated with the Dq brane probe are localized. Obviously $d_s \leq$ p
(the defect can at best fill all of the field theory dimensions). In order to be able to turn on an electric field so we can talk about a conductivity we also need $d_s \geq 1$,
the heat capacity can be calculated even for $d_s=0$.
For massless flavors, the only field that is turned on is the time component of the gauge field; it is governed by the DBI action
\begin{equation}
\label{DBI_0}
S_{DBI} = -  \N \, \int
\, du\,  u^{\nu} \sqrt{1 - A_0'^2} \,,
\end{equation}
where
\begin{equation}
\label{nudef}
\nu= {(p-7) (q-2 d_s-4+p)\over 4}+q-d_s-1\,,
\end{equation}
and the prime denotes a derivative with respect to $u$. The prefactor $\N$ is the product of the brane tension and the volume of the internal manifold and so depends on details of the probe brane. Fortunately we will not need it here. We also absorbed a factor of $2\pi \alpha'$ into $A_0$. As $A_0$ only appears derivatively, we can directly integrate the equations of motion
\begin{equation}
\label{sol}
A_0' = \frac{\tilde d}{\sqrt{ u^{2\nu} + \tilde{d}^2}} = 1 - \frac{1}{2} \frac{u^{2\nu} }{\tilde{d}^2} + {\cal O}(u^{4\nu})\,,
\end{equation}
where the integration constant $\tilde{d}$ as in \cite{Karch:2008fa} is proportional to the baryon number density $d$,
$\tilde{d} = (2 \pi \alpha' {\cal N} )^{-1} d$. Note that both gauge field solution and on-shell lagrangian are
actually independent of temperature. The temperature only comes into play as the lower end of the integration region.

The on-shell value of the action divided by the volume of the field theory spacetime gives us minus the free energy density $\omega$ in the grand-canonical ensemble. Plugging back in the solution for $A_0'$ this gives
\begin{equation}
\label{freeenergy}
\omega = {\cal N} \int_{u_H}^{\infty} \, du \,  \frac{u^{2\nu}}{\sqrt{u^{2\nu} + \tilde{d}^2}}
\end{equation}
where the divergent integral in $\Omega$ can simply be regulated by background subtraction or a local counterterm.
Since this is the free energy in the grand-canonical ensemble, we want to think of $\tilde{d}$ being a function of $T$ and $\mu$, where $\mu$ is the baryon number chemical potential. They are related via the condition that $A_0$ vanishes at the horizon, while $A_0=\mu$ at infinity by the standard AdS/CFT dictionary. Hence
\begin{equation}
\label{mudrel}
\mu = \int_{u_H}^{\infty} \, du \,A_0'.
\end{equation}
At zero temperature this gives \beq \tilde{d}= \gamma  \mu^{\nu} \eeq where
$\gamma = \left ( \frac{1}{\sqrt{\pi}} \Gamma (1
 +\frac{1}{2\nu} ) \Gamma(\frac{1}{2} - \frac{1}{2\nu} )\right )^{-\nu}
$
and
\beq
\label{eqofstate}
\omega=-\frac{\gamma {\cal N}}{ 1+\nu} \left ( \frac{\tilde{d}}{\gamma }\right )^{1+\frac{1}{\nu}}.
\eeq
At finite temperatures the integrals can still be done explicitly in terms of incomplete Beta functions \cite{Karch:2007br}. This was used in \cite{Karch:2008fa} to calculate the specific heat for the D3/D7 system and in \cite{Kulaxizi:2008jx} for the D4/D8 system.
But to extract the low temperature behavior of the free energy and the specific heat, it is sufficient to directly expand the integrands in eqs. (\ref{freeenergy}), (\ref{mudrel}) at small $u$ to extract the leading contributions at small $u_h$:
\begin{eqnarray}
\omega &=& {\cal N} \int_0^{\infty} {\cal L} - {\cal N} \int_0^{u_H} {\cal L} =- \frac{\cal N}{1+\nu} \gamma
\left ( \frac{\tilde{d}}{\gamma } \right )^{1+\frac{1}{\nu}} + {\cal O}(u_h^{2\nu +1}) \nonumber \\
\mu &=& \int_0^{\infty} A_t' - \int_0^{u_h} A_t' = \left ( \frac{\tilde{d}}{\gamma  } \right )^{\frac{1}{\nu}}  - u_h +
{\cal O}(u_h^{2\nu +1}) \nonumber \\
\left ( \frac{\tilde{d}}{\gamma } \right )^{\frac{1}{\nu}}&=& (\mu + u_h) + {\cal O}(u_h^{2\nu +1}).
\end{eqnarray}
As in \cite{Karch:2007br} we neglected density independent terms coming from the regulator. They have the same temperature dependence as the leading order $N_c^2$ term from the adjoint matter. Since they do not
depend on the density at all they should not be regarded as a contribution from the fluid we are studying, but simply a correction to the energy of the background plasma due to the presence of the fluctuations of
the dynamical flavor fields.
The corresponding density dependent part of the entropy density to leading order in the temperature is
\begin{equation}
\label{entropy}
s_{fluid} = - \left . \frac{ \partial \omega}{\partial T} \right |_{\mbox{fixed } \mu }=
  {\cal N} \, \tilde{d} \, \frac{\partial u_h} {\partial T} = d \, \frac{\partial\Delta m}{\partial T}.
\end{equation}
So as claimed, the entropy density is just $d$ times the entropy of a single quark. The same would be true for the free energy in the canonical ensemble, which we can obtain from the free energy $\omega$ of the grand canonical ensemble via a Legendre transform.
Finally, the specific heat $c_V$ at constant volume and chemical potential is (for $p \neq 3$)
\begin{equation}
\label{cv}
c_V = T \left . \frac{\partial S}{\partial T} \right |_{\mbox{fixed } d}= d \,T \, \frac{\partial^2 (\Delta m)}{\partial T^2} \sim
T^{\frac{2}{5-p}}.
\end{equation}
For the special case of $p=3$ we have that $u_h = \pi T$ is linear in temperature, and hence so is the free energy of the single string or the probe brane. The leading contribution to the entropy is then a constant, temperature
independent term that survives even at zero temperature as emphasized in \cite{Karch:2007br}. It would be very
interesting to understand from the field theory point of view how a single heavy quark in ${\cal N}=4$ SYM can
have a zero-temperature entropy of order $\sqrt{\lambda}$. In any case, this leading constant contribution
to the energy doesn't make it into the heat capacity. Latter is then dominated by the first subleading term which then scales as $T^{2 d_s}$ as shown in \cite{Karch:2007br}.

Another interesting quantity one can compute is the speed of sound at low temperature.
Using the equation of state (\ref{eqofstate}) one can obtain the following expressions for
the pressure and energy density
\begin{eqnarray}
\label{eosep}
   P &=& \frac{\cal N}{1+\nu} \gamma \mu^{\nu+1} \nonumber \\
   \epsilon &=& \nu P
\end{eqnarray}
which implies $c_s^2=1/\nu$. For D3/D7 system we recover conformal
value $c_s^2=1/3$, while for D4/D8 we get $c_s^2=2/5$ which is larger then
the upper bound on the speed of sound proposed in \cite{Cherman:2009tw,Hohler:2009tv}. Unlike the specific heat (and the conductivity which we discuss below)
the speed of sound in general depends on all 3 integers characterizing
the embedding, p, q and $d_s$. In Figure (1) we display the
speed of sounds for various Dp/Dq systems with 4 ND and 6 ND directions
respectively. Interestingly, for p=4 the speed of sound is always above
the conformal value and for p=2 below.
\begin{figure}
\label{csplot}
\begin{center}
\includegraphics[scale=.83]{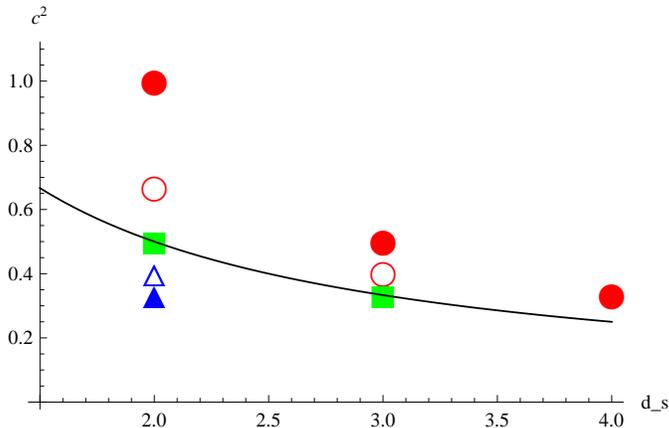}
\end{center}
\vspace*{-10pt}
\caption
  {Speed of sound squared $c^2$, for all Dp/Dq systems
with p=2, 3 or 4 and 4 or 6 ND directions, as a function of
$d_s$, the number of spatial dimensions. Shape/color
distinguishes p=4 (red circles), p=3 (green squares) and p=2 (blue triangles).
Open/filled symbols are 6/4 ND direction configurations. For p=3 the open
and filled squares lie on top of each other.
}
\vspace*{-10pt}
\end{figure}

\section{Massive quarks and specific heat}

In the previous section we showed that the low temperature behavior of the entropy
of a Dq probe brane describing massless flavors, is equal to entropy of a single
quark times the number density of the quarks $d$. To understand the origin of this
result, we will consider here a slightly more complicated case. We will compute
the free energy and specific heat of a Dq probe with a non-trivial embedding
profile, describing massive quarks in the dual field theory.

Giving mass to the field theory quarks, requires a Dp/Dq brane setup with
space transverse to both sets of branes.  The embedding of the Dq brane, is then
parametrized by one of the angular coordinates of this transverse space.
Writing the metric of the $(8-p)$-th dimensional sphere in the form
\beq
\label{sphereseparation}
d\Omega_{8-p}^2=d\theta^2+\sin^2\theta d\Omega_{k}^2+\cos^2\theta d\Omega_{8-p-k-1}^2
\eeq
defines the embedding function $\theta(u)$. As discussed before,
$\Omega_k$ represents the $k= \mbox{q} -d_s-1$ dimensional sphere wrapped by the Dq brane.
For reasons of stability we will only consider Dp/Dq systems which are
supersymmetric in the limit of zero temperature, thus $q=p+2,p+4$ and $k=2,3$
respectively\footnote{The case $q=p$ is special and will not be discussed here.}.

The induced metric on the Dq brane is
\beq
\label{Dqmetric}
ds_{Dq}^2=u^{\frac{7-p}{2}}\left(-f(u)dt^2+dx_{d_s}^2\right)+\frac{u^{-\frac{7-p}{2}}}{f(u)}\left[1+u^2 f(u)\left(\partial_u\theta\right)^2\right] du^2+u^{\frac{4-7+p}{2}}\sin^2\theta d\Omega_k^2
\eeq
and the DBI action for this configuration
\beq
\label{DBI_m}
S_{DBI}=-\N \int du \sin^k\theta u^{\nu}\sqrt{1+u^2 f(u)\left(\partial_u\theta\right)^2-A_0'^2}
\eeq
The prefactor $\N$ is the product of the brane tension, the time interval and the internal sphere $\Omega_k$.
$\nu$ is defined in (\ref{nudef}) and $A_0$ is rescaled by $2\pi\alpha'$ as in the previous section\footnote{
The Wess-Zumino term trivially vanishes for stable Dp/Dq systems with the proposed profile.}.
Observe that $A_0'$ appears only derivatively in the action so we can integrate its equation of motion
in this case, too.
\beq
\label{GaugeFieldm}
A_0'=\sqrt{\frac{1+f(u)u^2\left(\partial_u\theta\right)^2}{1+\frac{u^{2\nu}\sin^{2k}\theta}{\tilde{d}^2}}}
\eeq
The equation of motion for the profile function $\theta(u)$ is not analytically
solvable except in the limit of zero temperature \cite{Karch:2007br}. Fortunately, the details
of the embedding will not be necessary.

Evaluating the action on-shell gives us the grand canonical potential. Here however we choose to work
with the Legendre transform of the action which corresponds to the free energy at fixed charge density $d$.
\beq
\label{LegendreT}
\F=\tilde{d}\N \int_{u_h}^{\infty} du {\cal L} \qquad\qquad
{\cal L}=\sqrt{1+f(u)u^2\left(\partial_u\theta\right)^2}\sqrt{1+\frac{u^{2\nu}\sin^{2k}\theta}{\tilde{d}^2}}
\eeq
We immediately see the benefit of dealing with the canonical ensemble. If not for the factor
$\sqrt{1+\frac{u^{2\nu}\sin^{2k}\theta}{\tilde{d}^2}}$, the free energy would be equal to that of
a single static string configuration times the charge density $d$
(which is evident
by comparing to the Nambu-Goto action in the string frame metric). We will explore this similarity
further in the following.

We are interested in the low temperature behavior of the free energy.
Given that $T\sim u_h^{\frac{5-p}{2}}$ an expansion of the free energy of the Dq probe
around $u_h=0$ corresponds to a low temperature expansion for any background $Dp$ brane with $p<6$.
\beq
\label{TaylorF}
\F=\F(u_h)_{u_h=0}+\left(\frac{\partial \F}{\partial u_h}\right)_{u_h=0} u_h+{\cal O}(u_h^2)
\eeq
Clearly, the zeroth order term does not contribute to thermodynamic quantities such as the entropy or the specific heat.
To evaluate the linear term it is convenient to follow an approach similar to the previous section and
rewrite the integral in (\ref{LegendreT}) as
\beq
\label{Ltwoterms}
\F=\F_1-\F_2=d\int_0^{\infty}{\cal L}-d\int_{0}^{u_h}{\cal L}
\eeq
Taking the derivative of the second term with respect to $u_h$ and then the limit $u_h\rightarrow 0$
we arrive at
\beq
\label{Ftwo}
\frac{\partial \F_2}{\partial u_h}|_{u_h=0}=d
\eeq
This follows from the fact that $f(u=u_h)=0$ while $\partial_u\theta|_{u_h}$ is finite.
For the trivial embedding $\partial_u\theta=0$ the first term in (\ref{Ltwoterms}) is independent of
the temperature. Then (\ref{Ftwo}) gives the only linear contribution to the free energy and we
recover eq. (\ref{entropy}). When however, $\partial_u\theta\neq0$, the contribution of $\F_1$
to the linear term in the free energy expansion (\ref{TaylorF}) may be non-trivial. This is because
the profile function $\theta(u)$ will generically depend on $u_h$. To evaluate this term, we rewrite it
as
\beq
\label{Faderivative}
\frac{\partial \F_1}{\partial u_h}=\int_0^{\infty}du \left(\frac{\partial {\cal L}}{\partial f(u)} \left[\partial_{u_h}f(u)\right]+
\frac{\partial {\cal L}}{\partial \left(\partial_u \theta\right)} \left[\partial_{u_h}\partial_u \theta\right]+
\frac{\partial {\cal L}}{\partial \theta} \left[\partial_{u_h}\theta\right]\right)
\eeq
Observe that in the limit $u_h\rightarrow 0$ the first term in (\ref{Faderivative})
vanishes since $\lim_{u_h\rightarrow 0}\left[\frac{\partial f(u)}{\partial u_h}\right]=0$ for $p<6$ and the
integral is finite after regularization. Using the equations of motion for $\theta(u)$ and performing a
partial integration, we reduce (\ref{Faderivative}) to a boundary term. More precisely,
\beq
\label{lastep}
\F=\F(0)+d u_h-u_h\lim_{u_h\rightarrow 0}  \left[\left(d \frac{\partial {\cal L}}{\partial \left(\partial_u \theta\right)}-
\frac{\partial {\cal L}_{bg}}{\partial \left(\partial_u \theta\right)}\right) \left(\partial_{u_h}\theta\right)\right]_{u_h}^{\infty}+{\cal O}(u_h^2)
\eeq
Here the boundary term is regulated by background subtraction\footnote{Holographic renormalization
will in principle induce additional finite terms to the result. These terms however, are independent of the
charge density $d$.} with ${\cal L}_{bg}$ the $d\rightarrow 0$
limit of (\ref{LegendreT}).

For stable Dp/Dq embeddings of the type considered in this section $\theta(u)$ behaves close to the
boundary like \cite{Karch:2005os,Benincasa:2009,Mateos:2007vc}
\beq
\theta\simeq \frac{\pi}{2}+\frac{\hat{m}}{u^{\frac{2}{5-p}}}+\frac{\hat{c}}{u^{\frac{2 k}{5-p}}}+\cdots
\eeq
where $\hat{m}$ and $\hat{c}$ are proportional to the mass and the condensate of the dual theory
but independent of the temperature. This implies that
\beq
\lim_{u\rightarrow\infty}\partial_{u_h}\theta=0
\eeq
thus any contribution from the embedding profile $\theta(u)$ to the free energy at low temperatures comes from
the near horizon region.

Let us now Taylor expand $\theta(u)$ in the vicinity of the horizon as
\beq
\label{Taylortheta}
\theta\simeq \sum_{\ell=0}^{\infty}\left(u-u_h\right)^\ell a_\ell
\eeq
Using (\ref{Taylortheta}) it is easy to see that $\frac{\partial {\cal L}}{\partial \left(\partial_u \theta\right)}$
vanishes when evaluated at the horizon, whereas $(\partial_{u_h}\theta)$ is finite.
We deduce therefore that the free energy at low temperatures for massive Dq probes --- except for a
temperature independent term --- is given by the free energy of a single string times the charge density.

The picture is now clear. The low temperature behavior of the free energy requires us to zoom
in the near horizon region. In this regime, the Dq brane resembles a long narrow cylinder
emanating from the horizon; a spike. This spike represents a bundle of strings dissolved
in the Dq brane giving rise to the electric field on it \cite{Kobayashi:2006sb}. It is these strings
which give the leading (non-trivial) contribution to the free energy at low temperatures. As a result,
the entropy and the specific heat for massive flavors will be given by (\ref{entropy}) and (\ref{cv})
respectively. We see that the details of the embedding are irrelevant at low temperatures,
with the entropy being independent of the mass and the condensate of the dual theory.


\section{Drag on a single quark and DC conductivity of the Dq probe}

For general metric the drag has been obtained in \cite{Herzog:2006se} following the same logic as in the AdS$_{p+2}$ calculation of \cite{Herzog:2006gh,Gubser:2006bz}. For metrics of the general form
$$ds^2= -g_{tt} dt^2 + g_{xx} dx^2 + g_{uu} du^2$$
the magnitude of the drag force is given by
\beq
\label{drag}
F_{drag} = \frac{v}{2 \pi l_s^2} g_{xx} =\frac{v}{2 \pi l_s^2} (u^*)^{(7-p)/2}
\eeq
where we used that the metric component is to be evaluated at the special radial coordinate $u^*$ where
$g_{tt} = v^2 g_{xx}$. For us this yields
\beq
u^* = \frac{u_h}{(1-v^2)^{\frac{1}{7-p}}}
\eeq
and so
\beq
F_{drag}  = v \left ( 2^{16-3p} \pi^{(13-3p)/2} \Gamma \left ( \frac{7-p}{2} \right )
(7-p)^{p-7}
 \right )^{\frac{1}{5-p}} \frac{\lambda^{\frac{1}{5-p}} T^{\frac{7-p}{5-p}}}{\sqrt{1-v^2}}
\eeq
For $p=3$ this reduces to $\frac{\pi}{2} \sqrt{2 \lambda} T^2$ again in agreement with \cite{Herzog:2006gh}.
The drag coefficient $\mu M$ of \cite{Herzog:2006gh} can still be introduced via the definition
\beq
 F_{drag}  = \mu M \frac{v}{\sqrt{1-v^2}}.
\eeq

This can easily be related to the leading density dependent term in the conductivity.
The full conductivity found in \cite{Karch:2007pd} has the form $\sigma_{full} =
\sqrt{\sigma_0^2 + \sigma^2}$, where $\sigma$ is a linear density dependent term and $\sigma_0$ is, at least for massless quarks, a density independent term due to the thermally populated quarks and antiquarks. Since $\sigma_0$ is density independent, we take $\sigma$, the leading density dependent term, as the definition of the conductivity of the finite density chunk of matter (which is immersed into a thermal bath with its own non-vanishing conductivity $\sigma_0$). Typically, $\sigma_0$ scales as a positive power of $T$ and so can be
neglected in the limit of high density,
low temperature in which we calculated the heat capacity in the last section.
In low dimensional defects sometimes $\sigma_0$ scales as an inverse power
of $T$ but in all cases but $p=4$ and $d_s=1$ $\sigma$ dominates
over $\sigma_0$ at low temperatures. For $p=4$, $d_s=1$ both
contributions to the conductivity scale as $T^3$ and so it
depends on the density which one dominates.
$\sigma_0$ can also be made arbitrarily small by increasing the mass \cite{Karch:2007pd}; taking $\sigma$ to dominate the conductivity this way makes it appear more natural that one is dominated by the properties of a single string.

According to \cite{Karch:2007pd} to leading order in the density $d$ the conductivity then is universally given by
\beq
\label{sigma}
\sigma = d (2 \pi l_s^2) g_{xx}^{-1}
\eeq
where this time $g_{xx}$ has to be evaluated at $u_*$ where $g_{tt} g_{xx} = (2 \pi l_s^2)^2 E^2$ and $E$ is the electric field. So for us
\beq
\label{uls}
u_* = u_h \left ( 1 + (2 \pi l_s^2)^2 E^2/u_h^{7-p} \right )^{\frac{1}{7-p}}
\eeq
and hence
\beq
\sigma^{-1} = \left ( 2^{16-3p} \pi^{(13-3p)/2} \Gamma \left ( \frac{7-p}{2} \right )
(7-p)^{p-7}
 \right )^{\frac{1}{5-p}} \, \lambda^{\frac{1}{5-p}} T^{\frac{7-p}{5-p}} \frac{{\sqrt{e^2+1}}}{d}
\eeq
where $e=(2 \pi l_s^2) E/u_h^{(7-p)/2}$. As in the D3/D7 example analyzed in \cite{Karch:2007pd} this leading
order $d$ resistivity directly follows from the drag force. A single quark experiencing the drag force of
eq.(\ref{drag}) reaches a steady state velocity $v_{steady}$ given by $F_{drag}= E$, or in other words
\beq
v_{steady} = \frac{E}{\mu M} \frac{1}{\sqrt{1 + E^2/(\mu M)^2}} = \frac{E}{\mu M} \frac{1}{\sqrt{1 + e^2}}.
\eeq
The resulting current for a finite density $d$ of such quarks is then simply $j_x = d v_{steady}$ yielding
precisely the conductivity eq.(\ref{sigma}). So indeed for any Dq probe brane in a given Dp background
the manifestly q-independent drag force for a single string gives rise to a universal (q-independent)
leading $d$ behavior of the conductivities.

\section{Resistivity at zero density}

So far we have been mostly focusing on the leading contribution to heat capacity and conductivity in the limit of large density, low temperatures. It is in this limit that we found universal (q-independent) properties. The
calculation of \cite{Karch:2007pd} also allows to determine $\sigma_0$, the conductivity of the thermal plasma itself, even without any density. As we mentioned above, the full conductivity simply adds the two contributions in quadrature. Unlike the leading density dependent $\sigma$, $\sigma_0$ does depend on details of the defect. In particular, it depends on the spatial dimension $d_s$ of the defect. Typically $\sigma_0$ is a positive power of temperature (or at least less negative than the leading density dependent piece) and so is irrelevant in the low temperature, large density limit. According to \cite{Karch:2007pd} $\sigma_0 \sim e^{-\Phi} g_{xx}^{(d_s-2)/2} g_{SS}^{k/2}$, again to be evaluated at $u_*$, where $u_* = u_h + {\cal O} (E^2)$ according to eq.(\ref{uls}).
$g_{SS}$ here denotes the prefactor of the metric components of the internal sphere, that is $ds^2 = \ldots +
g_{SS} d \Omega_k^2$. At $u=u_h$ we have that $e^{-\Phi} \sim H^{(p-3)/4} \sim u_h^{(p-7)(p-3)/4} \sim T^{(p-3)(p-7)/(2 (5-p))}$ whereas $g_{xx}^{1/2} \sim u_h^{-(p-7)/4} \sim T^{-(p-7)/(2(5-p))}$ and $g_{SS}^{1/2} \sim u_h^{(p-3)/4} \sim T^{(p-3)/(2(5-p))}$. With this it is straightforward to calculate the temperature dependence of the resistivities at zero density. In particular,
\beq
\label{resistivityx}
\rho_0\sim T^{x} \qquad\qquad x=-\frac{2}{5-p} \left[ \frac{(p-7)(q-2 d_s-2+p)}{4}+q-d_s-1 \right]
\eeq
For the D4/D8 system studied previously in \cite{Bergman:2008sg} $\rho_0\sim T^{-2}$. This is in agreement
with the result obtained from (\ref{resistivityx}) for $p=4, q=8$ and $d_s=3$.

\begin{figure}
\begin{center}
\includegraphics[scale=.83]{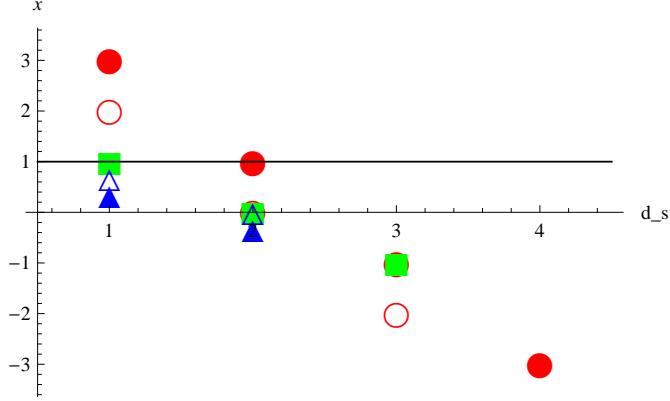}
\end{center}
\vspace*{-10pt}
\caption
  {Resistivity scaling factor $x$, for all Dp/Dq systems
with p=2, 3 or 4 and 4 or 6 ND directions, as a function of
$d_s$, the number of spatial dimensions. Shape/color
distinguishes p=4 (red circles), p=3 (green squares) and p=2 (blue triangles).
Open/filled symbols are 6/4 ND direction configurations.
}
\vspace*{-10pt}
\end{figure}

Note that among the various probe branes with 4 or 6 ND directions, whose resistivities are displayed in Figure (2), examples with resistivity linear in T as observed in the strange metal phase of high $T_c$ superconductors, are $p=3$, $d_s=1$ for any $q$ as well as $p=4$, $d_s=2$ with $q=4$. For the conformal $p=3$ case, $d_s=1$, that is we study a 1+1 d defect in a conformal theory. In this case linear resistivity is indeed guaranteed by conformal invariance. As current and charge densities have scaling dimension $d_s$ and the electric field has scaling dimension 2 in any dimension, the resistivity has dimension $2-d_s$, and in a conformal theory has to scale as $T^{2-d_s}$ as this is the only scale available. To see such a behavior in a higher dimensional (that is $d_s >1$) conformal theory with a gravitational dual, one would want to see an emergent AdS$_3$ geometry in the infrared, very similar to the emergent AdS$_2$ that has been seen in the recent studies of \cite{Faulkner:2009wj}.

\section{Conclusions}

To summarize, we observed that some thermodynamic and transport properties
of the Dp/Dq systems at finite density and low temperatures do not depend
on the dimensionality of the probe Dq brane.
In particular, the leading  behavior of the specific heat at low temperatures (summarized in
Table 1) is $q$-independent.
In the case of $p=4$ this implies a heat capacity linear in $T$ at low temperatures,
a characteristic behavior of Fermi liquids.
This is puzzling from the field theoretic point of view: while for $q=8$
the massless degrees of freedom coming from the D4/D8 intersections are fermions,
it is no longer the case in general, and charged bosons are expected to condense
at finite chemical potential.
We leave the detailed analysis of degrees of freedom  responsible for such a behavior
for future investigation.

The $p=3$ case is an exception, since the leading behavior of the specific
heat at low temperatures now depends on $q$.
One can trace this down to the fact that the leading, $q$-independent term in the entropy at low
temperature is constant and hence does not contribute to $c_V$.
The existence of a degenerate ground state is interesting in its own right:
for example this is a necessary feature of the models investigated in \cite{Faulkner:2009wj}
and might be responsible for the deviation from the Landau Fermi liquid behavior.

We also investigate the case of massive embeddings and find
that the  universality of the low temperature behavior of free
energy and specific  heat is a generic feature of the DBI action.
In this limit the leading contribution comes from the narrow long
tube embedding which can be also described as a bundle of fundamental strings.
The details of the embedding are irrelevant in this regime, and leading
order behavior is not sensitive to the mass and condensate of the dual theory.

Another interesting $q$-independent quantity is the leading (in charge density)
DC conductivity of the Dp/Dq brane system.
As in the case of specific heat, the $q$-independence is related to
the fact that the charge transport can be described in terms of
fundamental strings.
In the presence of boundary electric field, the equilibrium velocity of the string
is determined by the drag force; the resulting (manifestly $q$-independent) conductivity precisely agrees
with the leading behavior of the Dp/Dq conductivity computed from the DBI action.

We have also investigated some properties of the Dp/Dq  systems
which are less universal and depend on the details of the probe brane.
In particular, Figure 1 summarizes the behavior of the speed of sound
propagating on the defects [see also eq. (\ref{eosep})].
It is possible to have  values of $c^2$ both smaller and larger than
the conformal value $c^2=1/d_s$.
The low temperature behavior of the density-independent component of
the resistivity is displayed in Figure 2.
Linear temperature dependence (relevant for the ``strange metal''
phase above the superconducting dome) shows up in two cases; for $1+1$ dimensional defects
in a four-dimensional superconformal field theory where it is dictated by conformal invariance,
and for $2+1$-dimensional defects in a five-dimensional supersymmetric field theory.

\section*{Acknowledgments}
We would like to thank Chris Herzog for useful comments. Special thanks to
Oren Bergman for identifying and correcting an error in a previous version of this paper.
M.K. wishes to thank the organizers of the Simons Workshop in Mathematics and Physics where part
of this work was completed. The work of A.K. was supported in part by the U.S. Department of Energy
under Grant No.~DE-FG02-96ER40956. M.K. acknowledges support from NWO Spinoza Grant.

\bibliography{dpdq}
\bibliographystyle{JHEP}

\end{document}